# High temporal stability of supercurrents in $MgB_2$ materials

J. R. Thompson,[1,2] M. Paranthaman,[1] D. K. Christen,[1] K. D. Sorge,[2] H. J. Kim,[2] and J. G. Ossandon[3]

[1]Oak Ridge National Laboratory, PO Box 2008, Oak Ridge, Tennessee, USA 37837-6061
[2]Department of Physics, University of Tennessee, Knoxville, Tennessee, USA 37996-1200
[3]University of Talca, Curico, Chile

## Abstract
Fine grained polycrystalline samples of $MgB_2$ superconductor were synthesized from the elements to contain < 5 % of impurity phases, according to X-ray powder diffractometry. The superconductive transition was sharp with a midpoint $T_c$ = 38.5 K. The magnetization in the vortex state was studied as a function of applied field *H*, temperature *T*, and time *t*. From the equilibrium magnetization, the London penetration depth was obtained. The supercurrent density $J(T,H,t)$ in the vortex state (derived from the irreversible magnetization) decreases approximately linearly with *T*, in contrast to the quasi-exponential falloff in high-$T_c$ superconductors. The current is highly stable in time, with normalized decay rates $S = -d \ln(J)/d \ln(t)$ well below those in high-$T_c$ materials. These results are compared with those of other superconductors.

## Introduction
The recent discovery [1] of superconductivity near 40 K in $MgB_2$ has stimulated intense interest in this compound. An observation of a large boron isotope effect [2] points to a central importance for electron-phonon coupling. Studies of the conduction of critical currents [3, 4] in polycrystalline materials in large magnetic fields are encouraging: the grains appear to be well linked, in stark contrast with high-$T_c$ cuprate superconductors (HTSC) with their vexing problem of weak-linkage between grains. Another prominent feature of the HTSC's is a rapid decay of (nearly) critical currents in the mixed state, which arises from easy movement of vortices ("giant flux creep") that are depinned either by thermal activation or by quantum tunneling [5]. Thus it is important to investigate the temporal stability of supercurrents in this new material with its markedly different structure and simple composition.

In this work, we investigate several mixed state properties of an array of isolated $MgB_2$ grains. Thus we obtain the London penetration depth, which we use to establish the basic scale of current density, the depairing current density. This we compare with the experimentally realized current density and explore its dependence on temperature and magnetic field. Finally, the normalized decay rate is presented, then compared with the rates in several other superconductors. We find that for a substantial range of field and temperature, the mixed state supercurrent density is highly stable in time, in contrast with most HTSC materials.

## Experimental Aspects
The materials were prepared using high purity elemental starting materials. Fine grained polycrystalline samples of superconductor were synthesized by reaction of the elements at 890 °C for 2 h in a crimped Ta tube, contained within an evacuated quartz ampoule. The resulting materials contained < 5 % of impurity phases per X-ray powder diffractometry. In the form of a porous sintered pellet, the material exhibited a sharp superconductive transition with a midpoint $T_c$ = 38.5 K. The shielding was complete, as measured in an applied field of 0.4 mT. For studies in the mixed state, the sample was powdered and dispersed in a clear epoxy (Stycast 1267) with a volume fraction of ~ 2 %. This procedure reduces the magnitude of the irreversible magnetization relative to the equilibrium component, as this ratio is proportional to $<d^4>/<d^3> \sim d$, where $d$ is a particle's transverse dimension. The average particle dimension was determined by optical measurement of particles contained in thin sections cut from the sample, yielding the value $<d^4>^{1/4}$ = 12 µm. The total volume of superconductor was obtained from its mass (6.6 mg) and the X-ray density of 2.7 g/cm$^3$. In addition to making the equilibrium signal more visible, powdering of the sample also insures that the critical currents flow in a better defined, intragranular geometry; this procedures removes the uncertainty about the size of effective current path that is inherent in studies of low density, porous materials.

The materials were studied magnetically in a SQUID-based magnetometer, at temperatures $T$ = 5–300 K, in applied magnetic fields $\mu_0 H$ up to 6.5 T. In the vortex state, we determined the volume magnetization $M(T,H,t)$ as function of $H$, $T$, and time $t$. (To convert $M$ in Gauss into units of A m$^{-1}$, multiply by $10^3$.) The equilibrium magnetization $M_{eq}$ was obtained by averaging the measurements in increasing and decreasing field, as illustrated below; in all cases, the background moment from the epoxy was removed by subtracting the temperature-independent magnetic moment measured at 50 K. The same hysteresis loops provide the supercurrent density $J = \Delta M/r$, where $\Delta M$ is the hysteretic difference in $M$ (in G) measured in increasing and decreasing field and where $r = <d^4>^{1/4}/2$, (in cm). For studies of the time dependence of $J$, the temperature was first stabilized and the sample was taken into the critical state by setting a large negative field; then the field was increased to 0.25 T, where the magnetic moment was measured versus time for a period of ~ 1 h; following this, the field was increased for similar measurements at 0.5, 1, and 2 T. After increasing the field to 4 T, a short series of measurements were conducted at the same fields in the decreasing branch of the hysteresis loop, to obtain *in situ* values of the average (equilibrium) moment.

**Experimental Results and Discussion**

The magnetization of small, isolated grains of $MgB_2$ was studied in the superconductive state, as described above. Two examples of these data are shown in Fig. 1, which plots $M$ at temperatures of 25 and 30 K versus applied field $\mu_0 H$ on a log scale. The open symbols show measured values and illustrate the hysteretic magnetization in the mixed state. First we consider the equilibrium magnetization obtained by averaging the up-field and down-field values. The results are denoted by the solid symbols in Fig. 1. Now, in ordinary London theory, $M_{eq}$ has a simple logarithmic field dependence, [6] for fields $H_{c1} << H << H_{c2}$, with

$$M(H,T) = -\varphi_0 \left(32\pi^2 \lambda^2 (T)\right)^{-1} \times \ln\left(\eta H_{c2} / B\right) \qquad (1)$$

Here $\varphi_0$ is the flux quantum, $\lambda$ is the London penetration depth, and $\eta$ is a constant of order unity. The semilog plot of $M$ versus $\mu_0 H$ in Fig. 1 shows that this widely used relation provides a good description of the data for a significant field range, as shown by the solid lines that are fitted to Eq. 1. From the slopes of these and similar fits, we obtain values for $\lambda(T)$.

The results for the temperature dependence of the London penetration depth are shown in Fig. 2 as a plot of $1/\lambda^2(T)$. According to Ginzburg-Landau theory, one has $1/\lambda^2(T) \propto (T_c - T)$ near $T_c$. This relation is illustrated by the dotted line in the figure, which extrapolates to zero at the same $T_c$ as measured in low field. Thus it is evident that the data are well behaved. The downward curvature at lower temperatures is expected. Fitting these data to the BCS temperature-dependence in the clean limit yields the curve (solid line) shown in Fig. 2, with $\lambda(0)$ = 110 nm. This dependence provides a better description than the BCS dirty limit, two fluid, or strong coupling dependencies. Due to increasing hysteresis of the sample, the data become less reliable for $T \leq 20$ K and were excluded from the fit. Assuming other temperature dependencies provides somewhat different extrapolated values, of course, as illustrated by the simple parabolic fit (dotted wave) in Fig. 2.

Next we consider the persistent current density $J$ in the mixed state. Obtained from the magnetic hysteresis $\Delta M$, results for $J$ are shown in Fig. 3 as a function of temperature for several fixed fields. Several features are noteworthy. First, the temperature dependence is rather simple, with $J$ falling off almost linearly as $T$ increases. This contrasts sharply with the typical behavior of $J$ in high-$T_c$ superconductors, which often exhibit a quasi-exponential falloff associated with rapid thermally activated flux creep. A second feature is the relatively high scale of the current density, multiples of $10^{10}$ A m$^{-2}$ at low temperatures. This is significantly higher than the values often reported for bulk ceramic materials, where porosity and possibly secondary phases may interfere with the intergrain transport of current. To put the present values into context, let us consider the depairing current density $J_D = c\varphi_0\left(12\sqrt{3}\pi^2\lambda^2\xi\right)^{-1} \approx 170\cdot10^{10}$ A m$^{-2}$; here we use the above result for $\lambda$ and the value $\xi$ =5.2 nm from Finnemore *et al* [7]. Relative to $J_D$, the experimental current densities in Fig. 3 are entirely reasonable, with $J$ being approximately 5 % of the depairing current density. Thus one may expect realized current densities to approach or exceed $10\cdot10^{10}$ A m$^{-2}$ with improved synthesis of thin films and bulk materials, both in transport and magnetic studies.

With increasing temperature and/or magnetic field, the current density decreases and eventually becomes immeasurable. Operationally, this defines an irreversibility line $B_{irr}(T)$ which is shown in Fig. 4. These values were obtained from data like that in Fig. 3, using a criterion of $3\cdot10^7$ A m$^{-2}$, at which $J$ has deteriorated by three orders-of-magnitude from its low temperature level. Again a simple, almost linear dependence is found, which is similar in magnitude and form to the $B_{irr}(T)$ reported earlier.[7]

Now we turn to the temporal stability of the persistent currents. A convenient measure is the (dimensionless) normalized decay rate $S = -\mathrm{d}\ln(J)/\mathrm{d}\ln(t)$, obtained from the time-dependent measurements described above. In Fig. 5 is presented $S$ versus $T$ in fields of 0.25–2 T. The decay at low temperatures is quite small, with values near 0.003. In terms of a power law relation between the electric field $E$ and current density $J$ with $E = E_0(J/J_0)^n$, the exponent $n$ is simply related to $S$ with $(n-1) = 1/S$. This would imply that $n \sim 300$, corresponding to very steep and sharp curves in transport I-V studies. With increasing temperature, thermally activated depinning of vortices becomes more probable and $S$ increases. Current decay typically accelerates as a system approaches the irreversibility line. The same feature is found here, as shown by the vertical arrows in Fig. 5 that mark respective irreversibility temperatures taken from Fig. 4. Just below these temperatures, $S$ is increasing rapidly, while above them the decay is so rapid that $J$ is unmeasurable on the time scale of the experiment. Bugoslavsky *et al* [8] have observed a similar rapid increase in flux creep as $MgB_2$ approaches the irreversibility line.

It is useful to provide some context for the low decay rates observed in magnesium diboride. Thus we consider $S$ as a function of reduced temperature $T/T_c$ for some representative superconductors, all in an applied field of 1 T. The results are shown in Fig. 6. Materials include a $V_3Si$ single crystal ($T_c$ = 16.6 K), tapes of $Bi_2Sr_2CaCu_2O_x$ both as-synthesized and containing randomly oriented columnar defects created by fission fragments induced by 0.8 GeV protons [9], and a single crystal of $YBa_2Cu_3O_7$ containing parallel columnar defects formed by 1 GeV Au-ions [10]. The striking feature is that the $MgB_2$ decay rate lies below all of these for nearly the entire range of reduced temperature. Compared with the HTSC materials, $S$ at low temperatures is smaller by a factor of 3–10, and lies below that of the A-15 phase crystal as well. While the relative positions of the materials in Fig. 6 will differ in differing magnetic fields, it is encouraging and noteworthy that the persistent current density in $MgB_2$ is highly stable in magnetic fields of significant technological importance.

In conclusion, studies of the reversible and irreversible magnetization of isolated $MgB_2$ particles show that the scale of current density in even as-synthesized materials is high, approximately 5 % of the depairing current density. In magnetic fields of technological relevance, the supercurrents are highly stable in time over a significant temperature range.

We thank A. D. Caplin for communicating their results prior to publication. Research was sponsored by the DOE Division of Materials Sciences, under contract DE-AC05-00OR22725 with the Oak Ridge National Laboratory, managed by UT-Battelle, LLC.

**Figure captions**

1. The magnetization $M$ of isolated particles of $MgB_2$ at temperatures of 25 K (squares) and 30 K (circles), plotted versus applied magnetic field on a log scale. Open symbols show experimental measurements made in increasing and decreasing field, while closed symbols denote the average, equilibrium magnetization. Solid lines show fits to standard London theory, Eq. 1.

2. The London penetration depth plotted as $1/\lambda^2$ versus temperature. The dashed line illustrates a Ginzburg-Landau dependence near $T_c$, while the solid line is a fit to BCS theory in the clean limit.

3. The temperature dependence of persistent current density $J$ for $MgB_2$ in applied fields $\mu_0 H$, as shown.

4. The irreversibility line $B_{irr}(T)$.

5. A plot of the normalized decay rate $S$ versus temperature, for $MgB_2$ in applied fields of 0.25, 0.5, 1, and 2 T. Dotted vertical arrows show the irreversibility temperature for each case.

6. The normalized decay rate $S$ versus reduced temperature $T/T_c$ for several superconductors in an applied field of 1 T. The $Bi_2Sr_2CaCu_2O_x$ materials were c-axis textured tapes, either as-synthesized or irradiated with 0.8 GeV protons to create randomly oriented columnar defects with area density ("matching field") $B_\varphi \sim 1.4$ T. The $YBa_2Cu_3O_7$ single crystal contains columnar defects with $B_\varphi =$ 4.7 T from 1 GeV Au-ion irradiation. For the HTSC materials, the field was applied parallel to the c-axis.

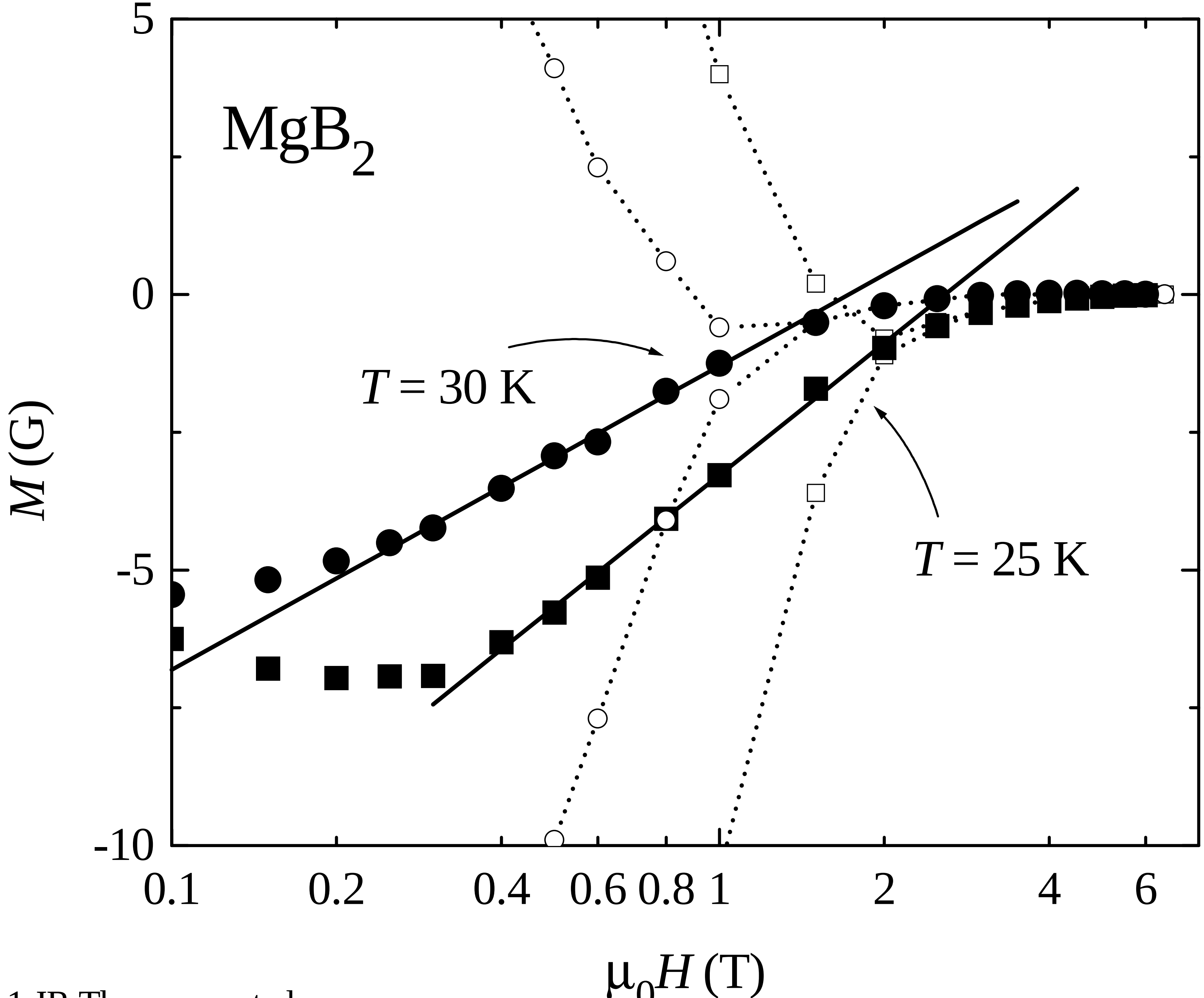

Fig. 1 JR Thompson et al.

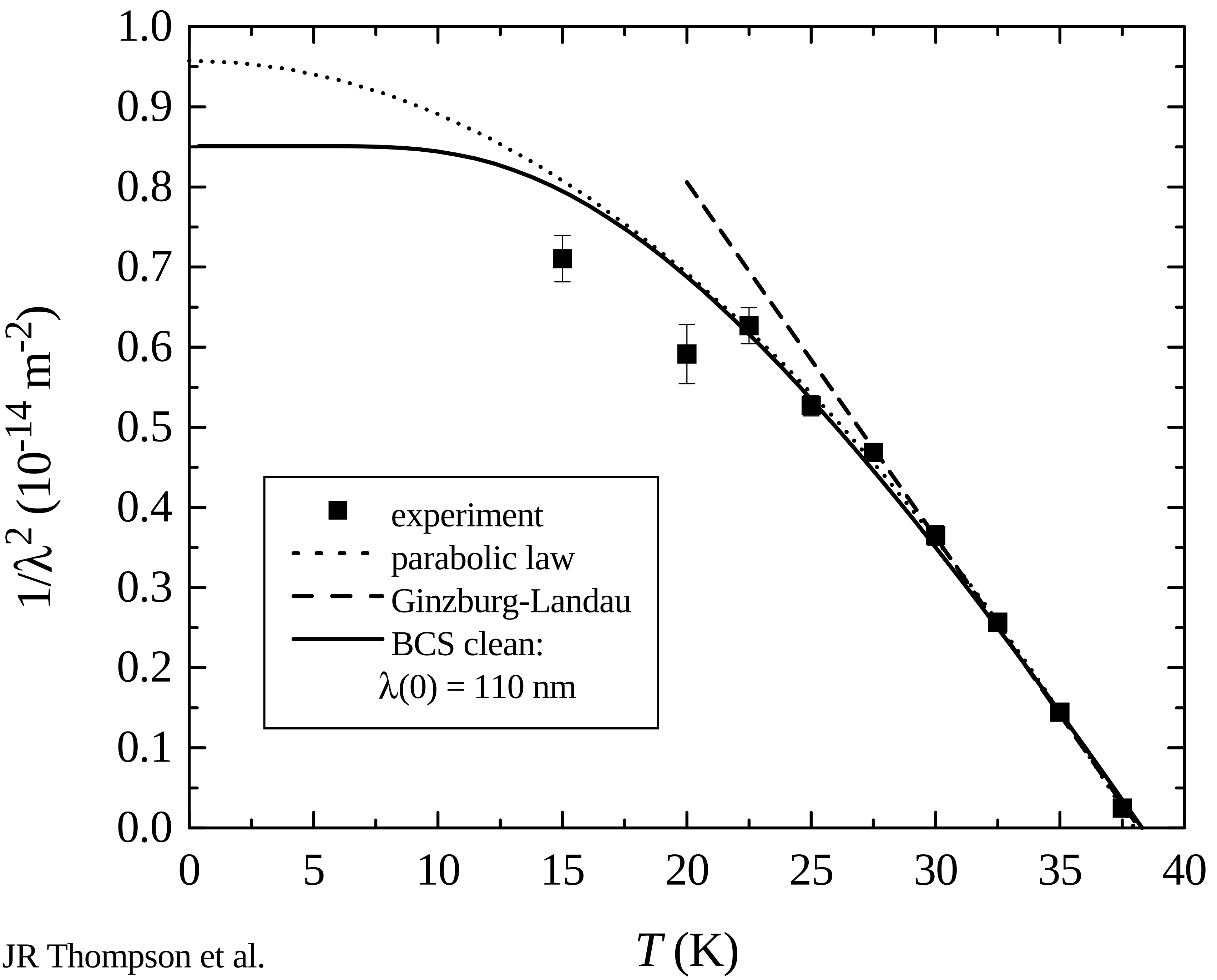

Fig. 2 JR Thompson et al.

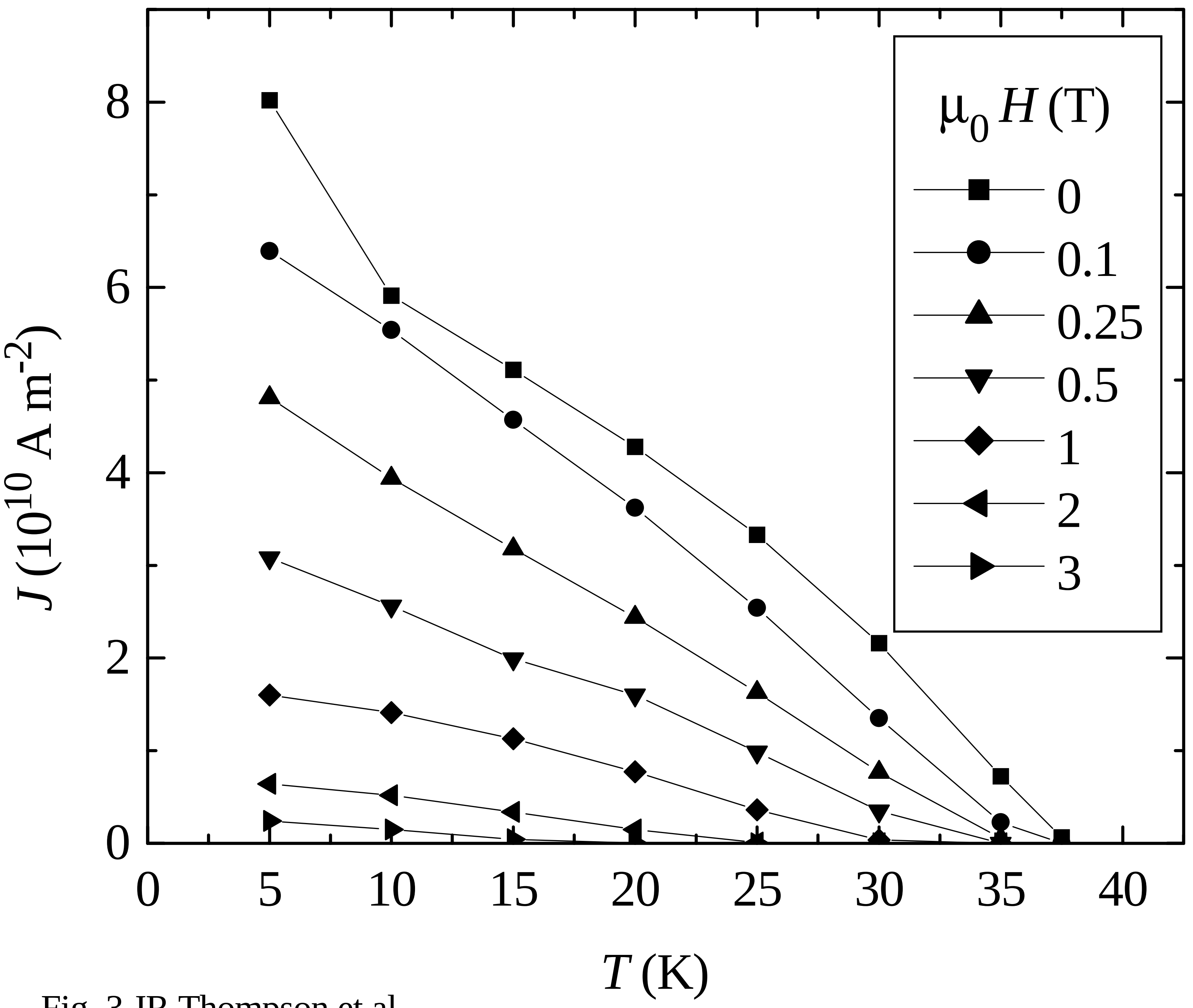

Fig. 3 JR Thompson et al.

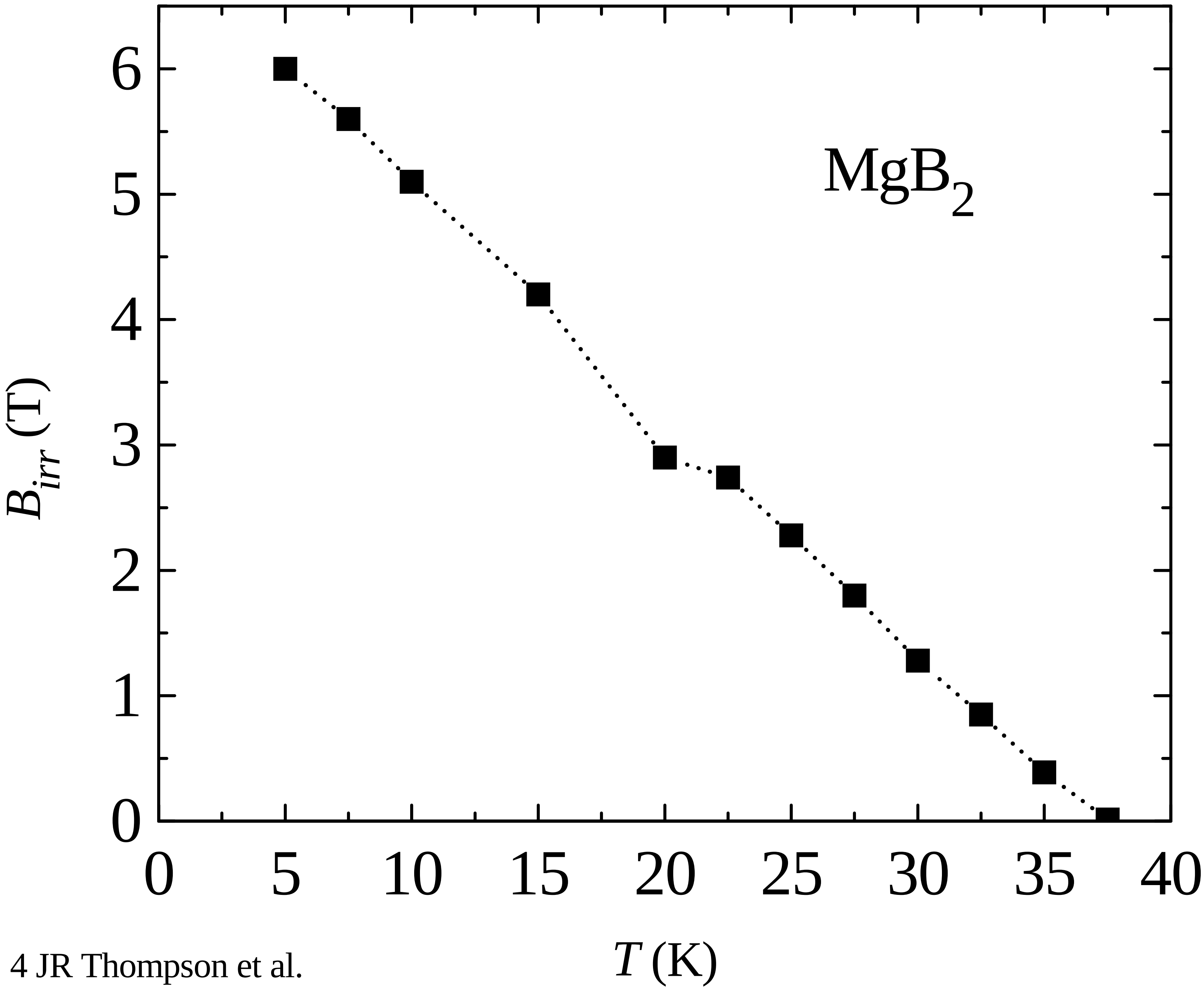

Fig. 4 JR Thompson et al.

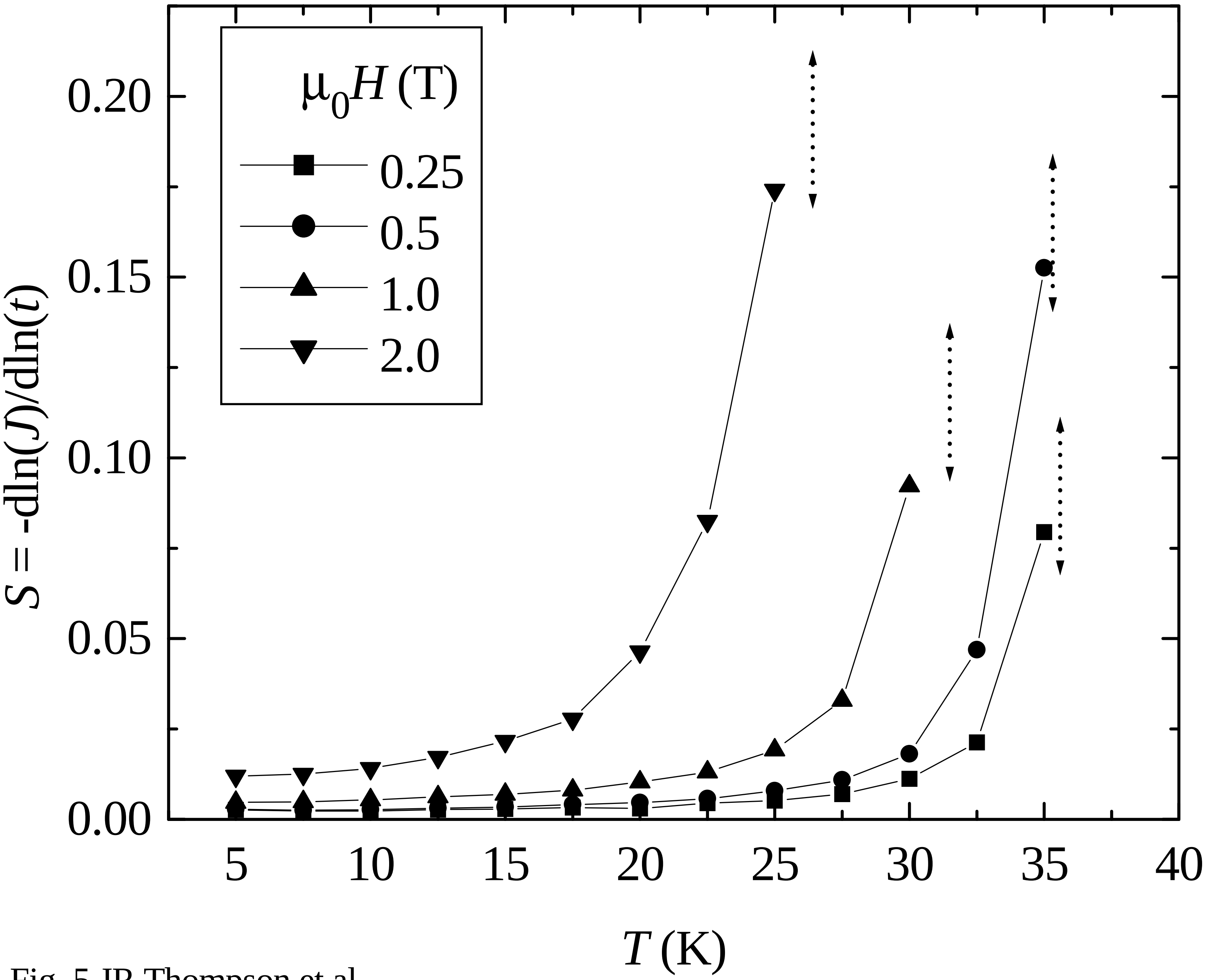

Fig. 5 JR Thompson et al.

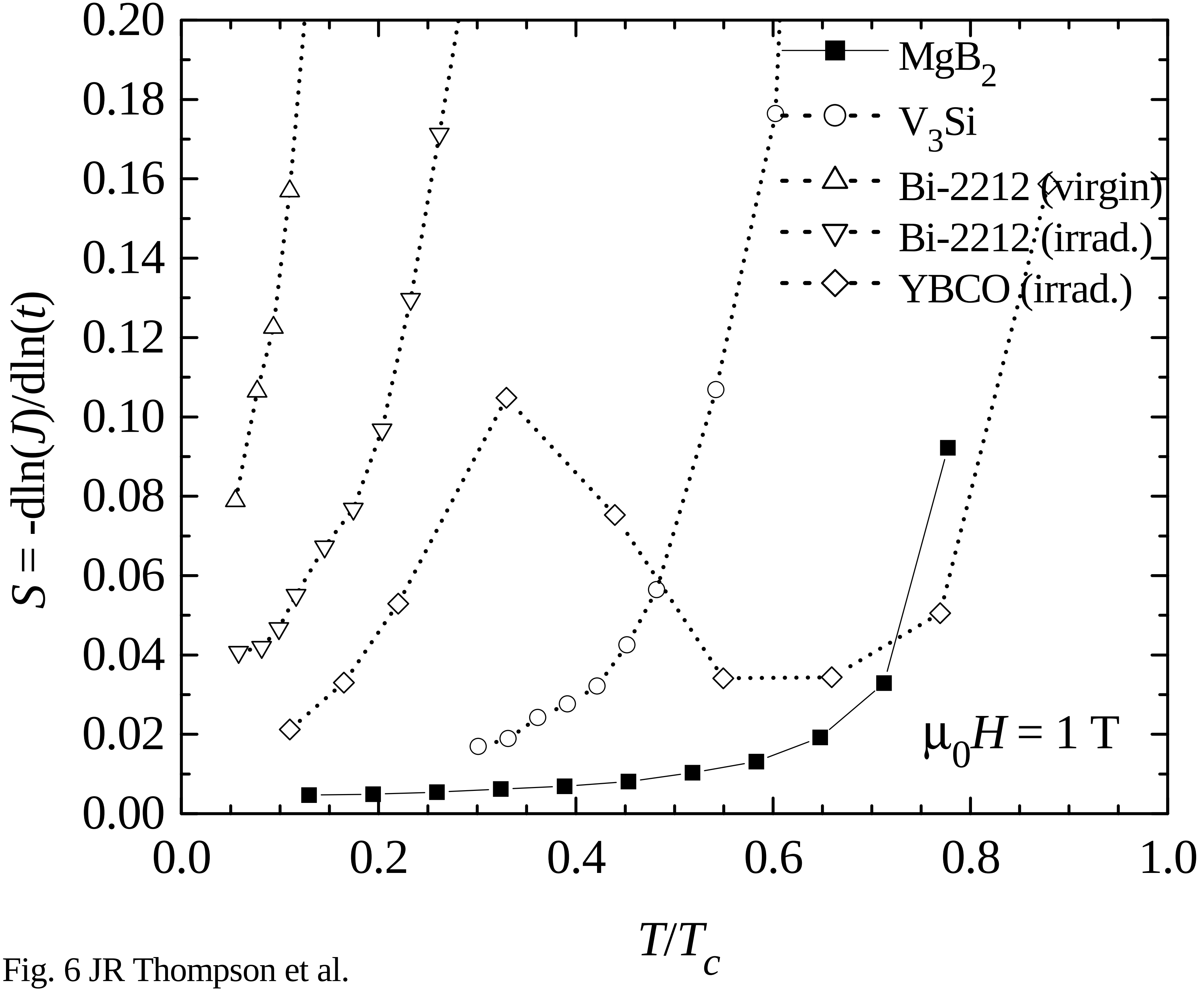

Fig. 6 JR Thompson et al.